\documentclass[%
 prl,twocolumn,
 nofootinbib,
groupedaddress,
 amsmath,amssymb,
 aps,
floatfix
]{revtex4-2}
\usepackage{amsmath}
\usepackage{graphicx,xcolor}% Include figure files
\usepackage{dcolumn}% Align table columns on decimal point
\usepackage{bm}% bold math
\usepackage{hyperref}% add hypertext capabilities
\begin{document}

%\preprint{APS/123-QED}

\title{Intrinsic limits on the detection of the anisotropies of the Stochastic Gravitational Wave Background}

\author{Giorgio Mentasti}%
\author{Carlo R. Contaldi}
\affiliation{%
 Blackett Laboratory, Imperial College London, SW7 2AZ, UK}
\author{Marco Peloso}
\affiliation{%
 Dipartimento di Fisica e Astronomia “Galileo Galilei” Universit\`a di Padova, 35131 Padova,
Italy \\
INFN, Sezione di Padova, 35131 Padova, Italy}
 
%\collaboration{MUSO Collaboration}%\noaffiliation

\date{\today}% It is always \today, today,
             %  but any date may be explicitly specified

\begin{abstract}
For any given network of detectors, and for any given integration time, even
in the idealized limit of negligible instrumental noise, the intrinsic time variation of the isotropic component of the Stochastic Gravitational Wave Background (SGWB) induces a limit on how accurately the anisotropies in the SGWB can be measured. We show here how this sample limit can be calculated and apply this to three separate configurations of ground--based detectors placed at existing and planned sites. Our results show that in the idealized, best--case scenario individual multipoles of the anisotropies at $\ell \leq 8$ can only be measured to $\sim 10^{-5} - 10^{-4}$ level over 5 years of observation as a fraction of the isotropic component. As the sensitivity improves as the square root of the observation time, this poses a very serious challenge for the measurement of the anisotropies of SGWB of cosmological origin, even in the case of idealised detectors with arbitrarily low instrumental noise.
\end{abstract}

\pacs{Valid PACS appear here}% PACS, the Physics and Astronomy
                             % Classification Scheme.
%\keywords{Suggested keywords}%Use showkeys class option if keyword
                              %display desired
\maketitle

{\sl Introduction.--} The field of Gravitational Wave (GW) detection has flourished in the recent years following the first observation of the signal from merging, massive, compact systems \cite{PhysRevLett.116.061102}. Many more detections have been made since the sensitivity of the existing LIGO--Virgo--KAGRA network continues to improve \cite{PhysRevD.104.022004}. The sensitivity of existing networks is expected to keep improving in the coming decades and new, ground--based detectors are already being planned \cite{LIGOisotropic2022,LIGOanisotropic2022}. While an increasingly large numbers of astrophysical signals arising from compact binary coalescence events have been observed, we are still awaiting the detection of a Stochastic Gravitational Wave Background (SGWB). The latest observations by LIGO--Virgo--KAGRA \cite{PhysRevD.104.022004} place upper limits $\Omega_{\rm GW} \lesssim 5.8\times 10^{-9}$ and $\Omega_{\rm GW} \lesssim 3.4\times 10^{-9}$ for, respectively, a scale-invariant and a stochastic signal from astrophysical sources with a power law spectral index of 2/3, pivoted at 25 Hz \cite{astrostochRegimbau,astrostochsiskalek}.

The SGWB is expected to be both of astrophysical and cosmological origin.  The astrophysical component originates from the superposition of a large number of unresolved sources, with the dominant contribution (at frequencies probed by ground--based detectors) from the coalescence of black holes and neutron star binaries. The cosmological component is more uncertain, with a possible origin from non-minimal models of inflation, phase transitions, and topological defects \cite{Caprini2018,LISAcosmogroup,Bartolo_2022}. 

The simplest way to disentangle these two components is through the frequency dependence of $\Omega_{\rm GW}$, the amplitude of the isotropic component of the SGWB \cite{Bartolo_2022}. Any anisotropy in the SGWB will also be of interest for the separation into different components. In particular, any directional dependence will assist in distinguishing between galactic and extra--galactic components and to search for any correlation with known tracers of structure \cite{SGWB-astro-sources,SGWB-astro-correl,SGWB-astro-correl-1,ValbusaDallArmi:2020ifo,Amalberti:2021kzh,ValbusaDallArmi:2022htu,Ricciardone:2021kel}. Anisotropies in the astrophysical background are expected to be correlated with the large scale structure, due to both how the GW originate and on how they propagate through a perturbed universe \cite{Contaldi:2016koz,Cusin:2017fwz,Bartolo:2019oiq}. Cosmological backgrounds, along with astrophysical ones in the limit of confused, persistent sources \cite{popcorn}, can be modelled as stationary phenomena.
%ones in the sense that their lagged correlation %function only depends on the time--lag. 
We will focus on this stationary limit in this work.

This {\sl letter} describes the application of a method developed for studying the Signal-to-Noise Ratio (SNR) of anisotropies in SGWBs as observed by a network of detectors \cite{Allen:1996gp}. We consider observations by a network of detectors placed at the location of existing (two LIGOs, Virgo and KAGRA  \cite{LIGO2,VIRGO2,KAGRA3}) and proposed (Einstein Telescope (ET) and Cosmic Explorer (CE) \cite{Maggiore:2019uih,CosmicExplorer}) instruments. Differently from the existing literature, here for the first time we investigate the ``noiseless limit'' of idealized instruments, where the sample variance of the observations dominates the error. Namely, we study the best-case scenario in which the noise of the instruments will be reduced to an arbitrarily low level. We find that, even in this ideal limit, it is impossible to achieve arbitrary resolution in the measurement of  anisotropies in the background because of the intrinsic variance of its {\sl isotropic} component. 

For simplicity, we consider an unpolarized and Gaussian SGWB, whose variance has a factorized dependence on the 
frequency $f$ and on the direction of observation $\hat n$ 
\begin{equation}
\left\langle h^*(f,\hat n)h(f,\hat n) \right\rangle=\frac{3H_0^2}{32\pi^3f^3}\, \Omega_{\rm{GW}}(f)\sum_{\ell m}\delta^{GW}_{\ell m}Y_{\ell m}(\hat n)\,,
\label{PSDdef}
\end{equation}
where $H_0$ is the Hubble constant and we choose to expand the frequency-independent anisotropies into spherical harmonic coefficients $\delta^{GW}_{\ell m}$ that are defined~\footnote{We verified that our definition is in agreement with that of  \cite{Bartolo_2022}.} relative to the energy density of the monopole, with $\delta^{GW}_{00}=\sqrt{4\pi}$ (so that $\Omega_{\rm{GW}}(f)$ is the standard fractional relative energy in GW per log frequency). We assume the anisotropy is much weaker than the monopole, i.e. $\vert \delta^{GW}_{\ell m} \vert \ll 1$, which is reasonable for both the cosmological and the astrophysical models, and that the $\hat z$-axis chosen to define the $Y_{\ell m}$ harmonic functions is aligned with the Earth rotation axis. 

{\sl Methodology --} We consider $N$ GW interferometers labeled by indices $i,j =1,2,\dots,N$. An SGWB results in a time-dependent signal $s_i(t)$, in each of the instruments, that depends on the response, the location, and the orientation of their arms (see e.g. \cite{Mentasti:2020yyd,Allen:1996gp}). The detectors are also affected by instrumental noise $n_i(t)$, such that each instrument results in a data stream
\begin{equation}\label{sandn}
m_i(t)=s_i(t)+n_i(t)\,.
\end{equation}

Following the formalism of \cite{Allen:1996gp,Mentasti:2020yyd} we Fourier-transform the data streams of detectors $i$ and $j$ over a time window $\tau$ centered at $t$ and build the observable $\mathcal{C}_{ij}(f,t)$ by cross-correlating them,
\begin{equation}
{\cal C}_{ij} \left(f,t \right) \equiv  { m}_{i}^\star \left( f ,\, t \right) { m}_{j}^{\,} \left( f ,\, t \right)\,,
\label{C-def}
\end{equation}
with $i\neq j$.
This is an unbiased estimator of the signal since the expectation value of the correlation of the noise in different sites vanishes. Taking advantage of the Earth rotation, we can compress the $\mathcal{C}_{ij}(f,t)$ into a set of functions $\mathcal{C}_{ij,m}(f)$ obtained through a finite-length, Fourier transform along the Earth rotation axis in a total observation time $T$
\begin{equation}
{\cal C}_{ij,m}(f)  \equiv  \frac{1}{T} \int_0^{T} dt \, {\rm e}^{-2\pi i\,  m  t/T_{e}} \, {\cal C}_{ij}  \left(f,t\right)\,,\label{C-m}
\end{equation}
where $T_e$ is the period of Earth rotation. We can integrate over frequency by applying a set of optimal filters $Q_{ij,m}(f)$, i.e. functions that can be chosen so as to maximize the signal--to--noise (SNR) of any given estimate. Namely, we define
\begin{align}
{\cal C}_{ij,m} \equiv\int_{-\infty}^\infty df\, {\cal C}_{ij,m}(f) \,Q_{ij,m}(f)\,.
\end{align}
For uncorrelated noise between different detectors, the expectation value $\langle\mathcal{C}_{ij}\rangle$ only contains contributions from the parameterized signal \eqref{PSDdef}, i.e.
\begin{equation}\label{CmExp}
\langle {\cal C}_{ij,m}\rangle\propto\int_{-\infty}^{\infty}df\,H(f)\sum_{\ell}\delta^{GW}_{\ell m}\gamma_{ij,\ell m}(f)Q_{ij,m}(f)\,,
\end{equation}
where $\gamma_{ij,\ell m}(f)$ are the response functions that depend on the geometry of the two detectors i and j, their distance vector and the multipole considered \cite{Allen:1996gp,Mentasti:2020yyd}.

For definiteness, we assume that the special shape of the signal is characterized by a constant $\Omega(f)=\Omega_0$, and
assume the fiducial value $\hat\Omega_0=10^{-9}$ in our explicit evaluations. However, due to the factorization \eqref{PSDdef} and the assumption of negligible 
instrumental noise, the accuracy in reconstructing the relative amplitudes $\delta^{GW}_{\ell m}$ is independent of the level and functional form of $\Omega(f)$. 
Furthermore, consistent with our goal of providing the best possible level with which any given multipole can be reconstructed, we assume that the 
signal is dominated by a monopole and a single anisotropic multipole, i.e.
$\left\langle h^*(f,\hat n)h(f,\hat n) \right\rangle=\Omega_{0}f^{-3}\left[1+\delta^{GW}_{\ell m}Y_{\ell m}(\hat n)\right]$\footnote{This means that we have to be careful when treating the search of a multipole $(\ell,m)$ with $m=0$. The analysis can be done also in this case, but the method is slightly different \cite{future}.}.
We consider the log-likelihood function $\ln \mathcal{L}( \Omega_0,\delta^{GW}_{\ell m})\sim -\chi^2( \Omega_0,\delta^{GW}_{\ell m})/2$ for the distribution of parameters relative to fiducial values $\hat\Omega_0$ and $\hat\delta^{GW}_{\ell m}$  with 
\begin{align}\label{likel_def}
&\chi^2=\sum_{m'm''}\sum_{\substack{i\neq j\\ k\neq l}} r^\star_{ij,m'} \Sigma_{ij,kl,m'm''}^{-2} r^{\,}_{kl,m''}\,,\nonumber\\
&r^{\,}_{ij,m'}( \Omega_0,\delta^{GW}_{\ell m})\equiv\mathcal{C}^{\,}_{ij,m'}\left(\Omega_0,\delta^{GW}_{\ell m}\right)-\left\langle\mathcal{C}^{\,}_{ij,m'}(\hat\Omega_0,\hat\delta^{GW}_{\ell m})\right\rangle,\nonumber\\
&\Sigma_{ij,kl,m'm''}^{2}=\left\langle r_{ij,m'}^\star
r^{\,}_{kl,m''}\right\rangle\,.
\end{align}
We assume a Gaussian Likelihood since the estimators $\mathcal{C}_{ij,m}$ are obtained by averaging over a large number of frequencies and solid angles, therefore we can assume that the central limit theorem holds. The assumption of small anisotropy, $\vert \delta^{GW}_{\ell m} \vert \ll 1$, further simplifies the computation of the covariance term in \eqref{likel_def}.
It can be shown \cite{future} that, after optimizing the filters $Q_{ij,m}(f)$, the chi-squared in (\ref{likel_def}) can be written, in compact form, as
\begin{equation}
\chi^2_{\rm opt}=T	\left[\left(\frac{\Omega_0}{\hat\Omega_0}-1\right)^2I_{00}+\frac{1}{4\pi}\left|\frac{\Omega_0}{\hat\Omega_0} \delta^{GW}_{\ell m}-\hat\delta^{GW}_{\ell m}\right|^2I_{\ell m}\right]\,,
\label{chisquared_opt_C}
\end{equation}
where the coefficients $I_{\ell m}$ are calculated from the integration over frequency of the response functions of the network of instruments over the variance term. In the signal dominated regime (i.e. when we can neglect the noise contribution and therefore set $n_i=0$ in \eqref{sandn}), the variance term just depends on the response functions of the combinations of the instruments in the network to the monopole. Therefore, in this limit the coefficients $I_{00}$ and $I_{\ell m}$ just depend on the geometry of the network and the orientation of the arms of the interferometers. This regime allows to determine what the ultimate limitations of a particular configuration of detectors will be in reconstructing the anisotropies of the SGWB.

We can now ask ourselves the question; {\sl what is the detection threshold for an anisotropy in the signal-dominated regime?}
Given the Gaussian assumption for the Likelihood we can forecast the expected error in a determination of parameters $\Omega_0$ and $\delta^{GW}_{\ell m}$ given their fiducial values $\hat\Omega_0$, $\hat\delta^{GW}_{\ell m}$. We do so by expanding the expression (\ref{chisquared_opt_C}) to quadratic order in the departure of the parameters from their fiducial values. namely, by computing the second derivatives of (\ref{chisquared_opt_C}), which form the so called Fisher matrix. %We observe that this matrix is diagonal in the two parameters $\Omega_0$ and $\delta_{GW,\ell m}$, leading to
%by considering the equivalent log-likelihood, $\mathcal{L}\sim -\chi^2_{\cal F}( \Omega_0,p_{\ell m})/2$, with
%\begin{align}\label{chisquared_fish}
%&\chi^2_{\cal F}=\begin{pmatrix} %\Omega_0-\hat\Omega_0\\ p_{\ell m}- \hat %p_{\ell m}
%	\end{pmatrix}^T
%	\cdot
%	\begin{pmatrix} %\mathcal{F}_{\Omega\Omega} & %\mathcal{F}_{\Omega p} \\ %\mathcal{F}_{p\Omega} & \mathcal{F}_{p p}
%	\end{pmatrix}\cdot
%	\begin{pmatrix} \Omega_0-\hat\Omega_0\\ %p_{\ell m}- \hat p_{\ell m}
%	\end{pmatrix}\,,
%\end{align}
%where
%\begin{equation}
%{\cal F}_{ab}=\left.\frac{\partial^2\chi^2_%{opt}}{\partial a\partial %b}\right|_{\substack{a=\hat a\\b=\hat %b}}\,.
%\label{fisherplm}
%\end{equation}
%
To estimate the detection threshold, we choose $\hat\delta^{GW}_{\ell m} = 0$, in which case the Fisher matrix is diagonal in the two parameters $\Omega_0$ and $\delta^{GW}_{\ell m}$:
\begin{equation}
\chi^2_{\cal F}=T	\left[\left(\frac{\Omega_0}{\hat\Omega_0}-1\right)^2I_{00}+\frac{1}{4\pi}\left|\delta^{GW}_{\ell m}\right|^2I_{\ell m}\right]\,. 
\label{chisquared_opt}
\end{equation}
%where we have chosen $\hat\delta_{GW,\ell m} = 0$.

The calculation of the integral measures $I_{\ell m}$ in \eqref{chisquared_opt}, in a general case, is complicated by the rotation of the network with respect to the sky, gaps in the data stream, and the presence of non-ideal noise features. In the signal-dominated regime, assuming $\vert \delta^{GW}_{\ell m} \vert \ll 1$, and aligning the coordinate frame (and thus the definition of $\delta^{GW}_{\ell m}$) with the Earth rotation, the calculation can be significantly simplified and carried out analytically \cite{future}.
Since we wish to consider the most idealized case, where we have a single anisotropic multipole $(\ell, m)$ on top of a dominant monopole, we have that the only non-vanishing expectation values of \eqref{CmExp} are $\langle\mathcal{C}_{ij,0}\rangle$ and $\langle\mathcal{C}_{ij,m}\rangle$. We note that in \eqref{chisquared_opt} the off-diagonal element of the Fisher matrix vanishes. This means that the observables $\mathcal{C}_{ij,m}$ with different values of $m$ are uncorrelated. This is due to the fact that we have arbitrarily chosen in \eqref{PSDdef} to expand the anisotropies along the Earth rotation axis.
The most general case, where one wishes to define the spherical harmonics expanded anisotropies with another axis would lead to non--diagonal terms and therefore an even larger sample variance. 
%For the sake of simplicity, we did not consider the study of an anisotropy with multipole $(\ell,0)$, since in that case the only non-vanishing expectation value of the observables is $\langle\mathcal{C}_{ij,0}\rangle$ and therefore a slightly different analysis must be performed \cite{future}.  
 
\begin{figure}[t!]
	\centering
	\includegraphics[width=\columnwidth]{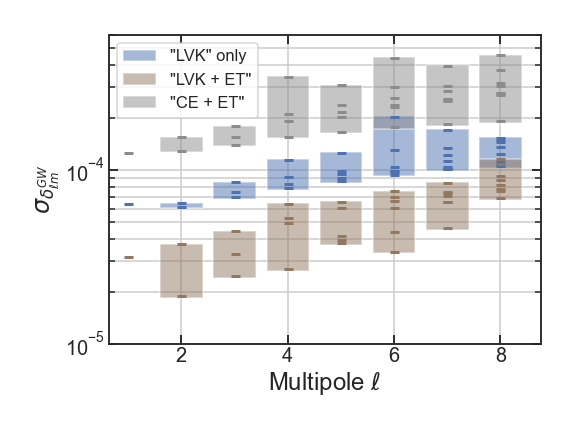}
	\caption{Estimated errors due to isotropic sample variance for multipoles $\delta^{GW}_{\ell m}$ at each $\ell \leq 8$. The shaded bands highlight the range of $\sigma_{\delta^{GW}_{\ell m}}$ at each $\ell$. The amplitude of the  $\delta^{GW}_{\ell m}$ is relative to the isotropic background $\hat\Omega_0$. The errors are marginalized over the background amplitude. Adding ``ET'' to the ``LVK'' network reduces sample variance at all multipoles, while ``CE+ET'' has worse sample variance.}
	\label{fig:plm_sigma}
\end{figure}

{\sl Results.--} We apply the estimate to three cases. The first is ``LVK'', a network consisting of instruments at the two LIGO, the Virgo, and the KAGRA sites \cite{LIGO2,VIRGO2,KAGRA3}.
The second case, ``LVK+ET'' adds a triangularly-shaped detector located at the proposed Sardinia site \cite{Maggiore:2019uih} for the Einstein Telescope. The third case, ``CE+ET'' consists of Cosmic Explorer and Einstein Telescope baselines only. We locate CE at the LIGO Hanford site. Consistently with the above discussion, we assume the signal-dominated regime across all detectors (namely, that the noise of all the instruments gives negligible contribution to the variances of the measurements). Throughout, we assume that data streams are Fourier--transformed on a short timescale $\tau=30$ seconds. This ensures that the sky rotation is negligible given the resolution of the detectors. However, the estimate is integrated over a much longer timescale $T=5$ years. We also assume a constant window in the frequency domain with $f\in [1,1000]$ Hz. Although the scenario used here is not band--limited by any noise filtering, this range mimics an optimistic frequency range of sensitivity for typical ground--based detectors. In practice, the effective window and bandwidth of the estimate will be determined by noise weighting. These assumptions, along with the coordinate alignment and noiseless limit mean our estimates constitute extreme {\sl best--case}, sample limited scenarios for each chosen configuration.

In Figure \ref{fig:plm_sigma} we show, for each network and for each multipole $\ell$ (up to and including $\ell=8$), the estimated standard deviation $\sigma_{\delta^{GW}_{\ell m}}\equiv \left(T I_{\ell m}/4\pi\right)^{-1/2}$ (as obtained after marginalizing \eqref{chisquared_opt} over $\Omega_0$), which provides an indication of the $SNR=1$ threshold for detection of the SGWB anisotropies. It is important to remember that the distribution of error amplitudes for a given multipole will change depending on the orientation of the coordinate frame, but the overall amplitude will be unchanged. In particular, the results show how the addition of baselines becomes important in the signal--dominated regime.

\begin{figure}[t!]
	\centering
	\includegraphics[width=\columnwidth]{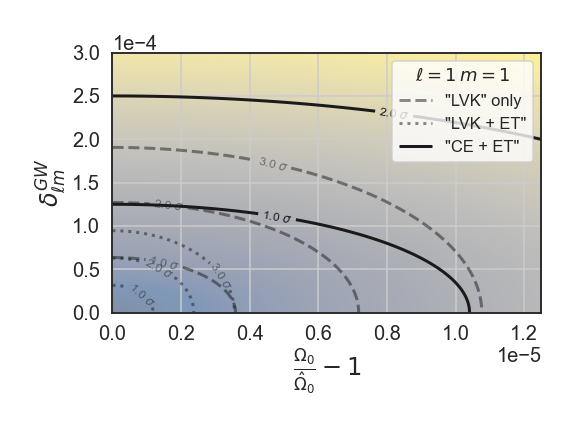}\\
	\caption{The $\chi^2_{\cal F}$ surface in both parameters for a single dipole mode. The $1\sigma$ to $3\sigma$ contours are highlighted for both ``LVK'' and ``LVK+ET'' networks. The $\Omega_0$ and $\delta^{GW}_{\ell m}$ parameters are uncorrelated for the choice of frame and variable definition adopted here. }
	\label{fig:fisher1}
\end{figure}

In Figure \ref{fig:fisher1} we show the function  $\chi_{\cal F}$ evaluated around the fiducial values in $\hat \Omega_0$ and $\hat\delta^{GW}_{\ell m}$ (for one example choice of $\ell$ and $m$). This also shows that the dominant effect, as one would expect in the absence of instrumental noise, is the number of baselines in the network.

%We stress that this result does not include any sensitivity limit since we assume a negligible contribution of the instrumental noise to the variance \eqref{C-def}. 

The results show that even in the extreme case of a future ground-based network with negligible noise, the SNR=1 threshold for the relative anisotropies is $\sim 10^{-5} - 10^{-4}$ over a period of measurement of ${\cal O}(10)$ years. This scale is determined by the effective bandwidth $\Delta f$ of the overlap functions, which is of the order of $\sim 100$Hz, and the total integration time $R$, which is of the order of $10^8$s. This crude estimate gives $\sigma \propto \frac{1}{\sqrt{T \, \Delta f}} \sim 10^{-5}$ reproduces well the scale of the sample variance, with an additional factor of $\sim 1-10$ (varying for the different networks considered) arising from the specific values of the function in the integrand of eq. (\ref{chisquared_opt}) \cite{future}. We stress that improving the $10^{-4}$ threshold by a factor of 10 requires increasing the measurement time by a factor of 100. Large angular scales anisotropies in the primordial background generated by the GW propagation in the early universe are expected to be below this level \cite{Starobinskii,Caprini2018}, with quadrupole amplitude $\sqrt{C_2}\sim 3\times 10^{-5}$ \cite{Bartolo_2022}. On the other hand, astrophysical backgrounds may have anisotropies above this level, and are expected to have background amplitudes within reach of future sensitivity \cite{Contaldi:2016koz,Cusin:2017fwz}.

Another signal to consider is the kinematic dipole of the SGWB induced by the Earth proper motion with respect to the background \cite{dipoleCMB}. We know, that our local system is moving with respect to the Cosmic Microwave Background (CMB) radiation rest frame \cite{10.1093/ptep/ptaa104} with a velocity $\beta \equiv v/c \sim 1.2\times 10^{-3}$ in the direction of galactic coordinates ($l = 264^\circ$, $b = 48^\circ$). This translates to an expected dipolar anisotropy with coefficients $\delta^{GW}_{1,-1}=-\delta^{GW}_{1,1}\simeq\sqrt{\frac{2\pi}{3}}\beta$ \cite{Allen:1996gp}. Since we have that, in the signal--dominated regime, $\sigma_{\delta^{GW}_{1,1}}=\sigma_{\delta^{GW}_{1,-1}}$, the combined measurement of the two multipoles allows us to have a final uncertainty on the kinematic dipole of $\sigma_{\beta}=\sqrt{\frac{3}{4\pi}}\sigma_{\delta^{GW}_{1,1}}$. This evaluates to $\sigma_{\beta}=3.4\times 10^{-5}$ for the LVK network and $\sigma_{\beta}=1.6\times 10^{-5}$ with the addition of ET. Therefore, sample variance is not an obstacle to detect the kinematic dipole induced by our peculiar motion, under the more conventional assumption that the  SGWB rest frame coincides with the CMB one.

{\sl Conclusions.--} We have examined the impact of sample variance on limiting best-case measurements of anisotropies in SGWBs, using networks of ground-based detectors. Our results show that different networks achieve SNR=1 at $10^{-5}-10^{-4}$ in the amplitudes (relative to the monopole) of the multiples of the SWGB. This constraint will need to be taken into account  next generation detector networks such as ``CE+ET'' (specifically, we show in \cite{future} that the expected sample variance is not much smaller than the design sensitivity of ET) and for a level of the isotropic signal that can be expected for the astrophysical backgrounds and for some cosmological scenarios \cite{Caprini2018}. 

%In that case, our results show that sample variance will be relevant for the characterization of anisotropies at which point the addition of baselines, with similar sensitivity, will yield the most improvement.  The outlook is better for astrophysical backgrounds and for the kinematic dipole, with the caveat that sensitivity improvements will still have to be made to reach detection thresholds. 

Our calculation disregards correlations between different multipoles in the most general case, where we expect more than a single multipole is present. We should expect these to be significant given the complicated and non-compact sky response functions of interferometer networks. This underscores that our estimates are to be considered as the most optimistic lower bounds on the sample variance of individual anisotropies. 

{\sl Acknowledgments.--} We thank Vuk Mandic for useful discussions. M.P. is supported by Istituto Nazionale di Fisica Nucleare (INFN) through the Theoretical Astroparticle Physics (TAsP) and the Inflation, Dark Matter and the Large-Scale Structure of the Universe (InDark) project. G.M. acknowledges support by the Imperial College London Schr\"odinger Scholarship scheme. C.R.C. acknowledges support under a UKRI Consolidated Grant ST/T000791/1.

%%%%%%%%%%%%%%%%%%%%%%%%%

\bibliographystyle{apsrev}
\bibliography{refs}
\end{document}